\documentclass{ws-ijmpe}
\usepackage[super,compress]{cite}
\usepackage{graphicx}
\usepackage{amsmath}
\usepackage{xcolor}
\usepackage{float}
\usepackage{amsmath}
\newcommand{\be}{\begin{equation}}
	\newcommand{\ee}{\end{equation}}
\newcommand{\ba}{\begin{eqnarray}}
	\newcommand{\ea}{\end{eqnarray}}

\begin{document}

\title{Study of Heavy Quarkonia in the presence of magnetic field by Nikiforov Uvarov method}

\author{Rishabh Sharma, Siddhartha Solanki, Manohar Lal and Vineet Kumar Agotiya*}

\address{Department of Physics, Central University of Jharkhand, Ranchi, India, 835 222.\\
	*Corresponding author id- agotiya81@gmail.com}

\maketitle
\begin{abstract}
The N-dimensional radial Schrodinger equation has been solved using the Nikiforov Uvarov (NU) method, in which we used the medium modified form of Cornell potential and quasi-particle Debye mass with strong magnetic field background. The binding energies and the mass spectra of heavy quarkonium have been studied in the N-dimensional space for different values of magnetic field, the binding energy decreases with increasing magnetic field, which shows early dissociation of heavy quarkonium system. The influence of dimensionality number has also been discussed on binding energies of $J/\psi$ and $\Upsilon$ for fixed value of magnetic field. It is found that with an increase in dimensionality, the binding energy starts decreasing from a higher initial value. The results obtained are quite consistent with recent studies.

\end{abstract}

%
%
%
%
%

\section{Introduction}
The study of heavy quarkonium system (charmonium and bottomonium) has played a crucial role in the understanding of the strong interaction i.e. quantum chromodynamics (QCD). Schrodinger equation is considered an important mathematical tool to deal with heavy quarkonium system in particle physics. For this reason, Schrodinger equation remains an interesting topic in the scientific community even today. One of the most challenging tasks in non-relativistic quantum mechanics ~\cite{1} is to find the exact and analytical solution for Schrodinger equation by some specific potentials of our physical interest. The exact solution of Schrodinger equation can be obtained for certain potentials. While considering inverse-power potentials~\cite{2,2a,3}, one has to obtain the approximate solution~\cite{4,5,6,7} using some approximations like the Pekeris approximation~\cite{7}.  Several methods to solve the Schrodinger equation have been designed from time to time as discussed in these studies~\cite{2,8,9,10,11}. The energy eigenvalue in the N-dimensional space for heavy quarkonium using the quadratic potential plus inverse of quadratic potential under the influence of finite temperature~\cite{12} and using medium modified form of Cornell potential in the presence of baryon chemical potential can be found in~\cite{12a}. 

Nikiforov Uvarov functional analysis (NUFA) method~\cite{15,16,17} is an extension of the NU method. NUFA method reduces some mathematical complexities, that are left in NU method. The limitation of this method is that it is valid only for specific exponential potentials. 
In recent years, the potentials which are used in analytical methods to find solutions in N- dimensions are Hulthen Hellmann potential~\cite{4,18,19} coulomb potential~\cite{20} Pseudo harmonic potential~\cite{21}, Mie type potential~\cite{22}, Hua potential~\cite{5} Screened Kratzer potential~\cite{23,24}, global potential~\cite{25}, and Cornell Potential with some modifications~\cite{26,27,28,29}. There are many other potentials that are being explored by the scientific community. Several methods such as asymptotic iteration method (AIM)~\cite{30,31,32,33,34}, Laplace transformation method~\cite{20,40}, supersymmetry quantum mechanics method (SUSYQM)~\cite{6,7,41,42}, power series method~\cite{43}, analytical exact iteration Method (AEIM)~\cite{44,50} has been used to get solution of Schrodinger equation in N-dimensions. NU method~\cite{3,4,13,13a,14} is also such a type of analytical tool which is widely used to obtained the solution of the Schrodinger equation. NU method is the most effective method to find a valid solution of Schrodinger equation to study various systems like diatomic molecules, poly-atomic molecules and heavy quarkonium. NU method is an elegant and analytical method based on 2nd order differential equation, by virtue of which we reduced the Schrodinger equation into hypergeometric form, further comparing the coefficients of generalized hypergeometric equation~\cite{14} and hypergeometric form of Schrodinger equation which pave us a way to calculate binding energy and wave function. It is important to mention here that this method is not only applicable for the exponential potential but also for non-central potential. A typical potential which has played a vital role in our understanding of the thermodynamic properties and energy eigenvalues of quarkonium is the medium modified form of Cornell potential~\cite{12a}. This potential is very suitable to find the spectral behavior of quarkonia in hot QCD medium. In the present work, the medium modified Cornell potential have been used to solve the Schrodinger equation using NU method and hence to study the quarkonium properties. The main reason for solving the Schrodinger equation with Cornell potential is that it is a central potential and this central potential has its important place almost in all branches of physics. In high energy and particle physics, central potentials are used to investigate the quarkonium interactions, in nuclear physics it describes the spectroscopic outcomes, in molecular physics these interactions are used to provide the intra-molecular and inter-molecular interactions etc. 

The medium modified Cornell potential has been already studied using SUSYQM method~\cite{12a}. 
Recent studies~\cite{46,47} focus on the effect of magnetic field and the transport properties of the heavy quarkonia this is because during the heavy ion collision a charged fermion during its motion produces magnetic field up to a scale of 0.3 $GeV^2$, which is nearly equal to the range of the magnetic field believed to exist in the early universe for a short time. Motivating from the above-mentioned study, we solve Schrodinger equation in N-dimensions. Here we solve the NU method for medium modified Cornell potential under the influence of background magnetic field, to examine the effect of magnetic field on the general behavior of binding energy and mass spectra of the heavy quarkonium. \\
Our present article is organized in the following manners:
In section \ref{sec2}, we provide a brief idea about NU method, medium modified Cornell potential and the magnetic field dependent Debye mass and the solution of Schrodinger equation has been obtained and in subsection \ref{subsec2.4}, we discuss the results and summary of the present work in section \ref{sec3}.


\normalsize

\section{Methodology} \label{sec2}

\subsection{Brief explanation of Nikiforov Uvarov method}
NU method~\cite{13} is an appropriate mathematical tool to get the solution of 2nd order differential equation by using coordinate transformation into hypergeometric type generalized equation~\cite{14} which can be written as,
\begin{equation}
	\label{1}
	\psi^{\prime\prime}\left(s\right)+\frac{\bar{\tau}\left(s\right)}{\sigma\left(s\right)}\psi^\prime\left(s\right)+\frac{\bar{\sigma}\left(s\right)}{\sigma^2\left(s\right)}\psi\left(s\right)=0.                 
\end{equation}

Where $\sigma(s)$ and $\bar{\sigma}(s)$ are the polynomials of second degree and $\bar\tau(s)$ is the polynomial of maximum first degree. To find the solution of Eq.~(\ref{1}), method of variable separable is used with appropriate coordinate transformation,
\begin{equation}
	\label{2}
	\psi(s) = \phi(s).\chi(s),                                                     
 \end{equation}

Eq.~(\ref{1}) is reduced to hypergeometric type equation in terms of $\chi(s)$. Here, $\chi(s)$=$\chi_n(s)$ is n degree polynomial to satisfy hypergeometric type equation:
\begin{equation}
	\label{3}
	\sigma(s)\chi''(s) + \tau(s)\chi'(s) + \lambda\chi(s) = 0,              
\end{equation}
 where 
 \begin{equation*}
 	\label{}
 	\sigma(s) =\pi(s) \frac{\phi(s)}{\phi'(s)},              
 \end{equation*}
 \begin{equation*}
 	\label{}
 	\tau(s)={\bar\tau(s)}+2\pi\left(s\right).	
 \end{equation*}

 $\pi\left(s\right)$ is a linear function of first degree and $\lambda$ is a parameter respectively written in Eq.~(\ref{4}) and Eq.~(\ref{5}) as below:  

\begin{equation}
		\label{4}
\pi\left(s\right)\ =\frac{\sigma^\prime\left(s\right)-\bar{\tau}\left(s\right)}{2}\pm\sqrt{\left(\frac{\sigma^\prime\left(s\right)-\bar{\tau}\left(s\right)}{2}\right)^2-\bar{\sigma}+k\sigma}(s).
\end{equation}

\begin{equation}
	\label{5}
	 \lambda=k+\pi^\prime\left(s\right).                                                      
\end{equation}
On the other hand, the value of  $k$  can be found by putting the discriminant equal to zero, and value inside the square root must be the square of the polynomial during the solution of Schrodinger equation.

In NU method to get the  eigen energy solution of Schrodinger equation, we have to used the quantization rule:
\begin{equation}
	\label{6}
	\lambda=\lambda_n=-n\tau^\prime\left(s\right)-\frac{n\left(n-1\right)}{2}\sigma^{\prime\prime}\left(s\right).
\end{equation}
After taking the values of $\sigma^{\prime\prime}\left(s\right)$, $\tau^\prime\left(s\right)$ and $\lambda$ from Eq.~(\ref{1}), Eq.~(\ref{3}) and Eq.~(\ref{5}) respectively, we get the final expression for the binding energy solution.

\subsection{Medium modified Cornell (MMC) potential}

Using the NU method, one can solve the Schrodinger equation for the quark-antiquark system where the underlying potential is of the Cornell type which has both the Coulombic and linear terms liable for the fundamental features of the QCD (i.e. interaction at the small distance and the confinement at the large distance). The study of this potential is of the particular interest for determining the binding energy of the coupled states and their mass spectra and hence to characterize the electromagnetic characteristics of the meson (heavy quark). The quarkonium interaction potential, which we consider in the present case is the medium modified form of Cornell potential \cite{12a,27,49,50} is given below: 
\begin{equation}
	\label{7}
	V\left(r\right)=\ \left(\frac{2\sigma}{m_D^2}-\alpha\right)\frac{e^{-m_D\ r}}{r}-\frac{2\sigma}{m_D^2\ r}+\frac{2\sigma}{m_D}-\alpha\ m_D.
\end{equation}
In Eq.~(\ref{7}), $\alpha$ is temperature dependent two loop running coupling constant the value of which is given below,
\begin{equation*}
	\label{}
	\alpha=\frac{6\pi}{\left(33-2N_f\right)\theta}\left(1-\frac{3\left(153-19N_f\right)}{\left(33-2N_f\right)^2}\frac{log2\theta}{\theta}\right).
\end{equation*}
Where, 
\begin{equation*}
	\label{}
	\theta=log\left(\frac{T}{\lambda_T}\right).
\end{equation*}
$\lambda_T$ and $N_f$ represents the normalized QCD coupling scale number of flavors respectively.  $\sigma$=0.224 $GeV^2$ is string term of interacting quark system~\cite{15a} and $m_D$ represent the magnetic field dependent quasi particle Debye mass.\\
On expanding Eq.~(\ref{7}) by Taylor series expansion, we get 
\begin{equation}
	\label{8}
	V\left(r\right)=\ -\left[\left(\frac{2\sigma}{m_D^2}-\alpha\right)\frac{m_D^3}{6}\right]r^2+\left[\left(\frac{2\sigma}{m_D^2}-\alpha\right)\frac{m_D^2}{2}\right]r+0r^0-\frac{\alpha}{r}.
\end{equation}
Rewrite Eq.~(\ref{8}) as follows:
\begin{equation}
	\label{9}
	V\left(r\right)=a_1r^2+a_2r+a_3r^0-\frac{a_4}{r}.
\end{equation}
Now comparing Eq.~(\ref{8}) and Eq.~(\ref{9}) we get:
\begin{equation*}
	\label{}
	a_1=-\left[\left(\frac{2\sigma}{m_D^2}-\alpha\right)\frac{m_D^3}{6}\right], \\
	a_2=\left[\left(\frac{2\sigma}{m_D^2}-\alpha\right)\frac{m_D^2}{2}\right], \\
	a_3=0,\\
	a_4=\alpha.  
\end{equation*}
Finally, the potential became of the form of
\begin{equation}
	\label{10}
	V\left(r\right)=a_1r^2+a_2r-\frac{a_4}{r}.
\end{equation}
Where the 1st term is quadratic used to improve the properties of quarkonium, linear shows the confinement feature and the 3rd term represent the Coulombic part. 
\subsection{Debye Mass in presence of strong magnetic field}
Debye mass $m_D$ is an important tool to understand the screening of color forces in hot QCD medium. It is known that the plasma shows collective behavior because both charged and neutral quasi-particles are found in it. Debye mass can be measured in terms of screening length (inverse of $m_D$) because of its ability to shield out the electric potential applied to it. Debye mass in the presence of strong magnetic field (eB) for hot QCD medium as found in \cite{46,51} is given by
\begin{equation}
	\label{11}
	m_D^2=4\pi\alpha\left(\frac{6T^2}{\pi}Polylog\left[2,z_g\right]+\frac{3eB}{\pi}\frac{z_g}{1+z_g}\right).     
\end{equation}
Here `eB' term is used for the representation of magnetic field, and the expression of Polylog function used in the above equation is given below,
\begin{equation*}
	\label{}
	Polylog[2,z_g]=\sum_{k=1}^{\infty}\frac{z^k}{k^2}.
\end{equation*}
$z_g$ represent the quasi gluon effective fugacity, For ideal Equation of states $z_g$ = 1, then Debye mass in terms of temperature and magnetic field \cite{46a} is given as
\begin{equation}
	\label{12}
    m_D^2\left(T,eB\right)=4\pi\alpha\left(T^2+\frac{3eB}{2\pi^2}\right).	
\end{equation}

\subsection{Approximate Solution of Schrodinger equation with MMC Potential} \label{subsec2.4}
According to quantum mechanics the two particle system can be described by Schrodinger equation \cite{11b}.The Schrodinger equation for two interacting quarks in N-dimensional space \cite{12,11a} can be written as:
\begin{equation}
\label{13}
\frac{d^2R}{dr^2}+2\mu\left[E-V\left(r\right)-\frac{\left[L+\left(\frac{N-1}{2}\right)\right]^2-\frac{1}{4}}{2\mu r^2}\right]R\left(r\right)=0.     
\end{equation}
Where `E' is the energy eigenvalue, N and L are the dimensionality number and angular momentum quantum number respectively. $\mu$ is the reduced mass of quarkonium system and ‘r’ is the inter-quark distance. Substituting the value of potential V(r) from Eq.~(\ref{10}) in Eq.~(\ref{13}), we get:
\begin{equation}
\label{14}
\frac{d^2R}{dr^2}+2\mu\left[E-a_1r^2-a_2r+\frac{a_4}{r}-\frac{\left[L+\left(\frac{N-1}{2}\right)\right]^2-\frac{1}{4}}{2\mu r^2}\right]R\left(r\right)=0.
\end{equation}
To make differentiation easier, we introduced an approximation scheme. Let us assume $r = 1/x$ then, differential equation modifies in the form of $x$ and we get on to differentiate it with respect to the ‘$x$’ as follows:
\begin{equation}
	\begin{aligned}
	\label{15}
	\frac{d^2R}{dx^2}+\frac{2x}{x^2}\frac{dR}{dx}+\frac{2\mu}{x^4}\left[E-\frac{a_1}{x^2}-\frac{a_2}{x}+a_4x-\left(\frac{\left[L+\left(\frac{N-1}{2}\right)\right]^2-\frac{1}{4}}{2\mu}\right)x^2\right]R(x)=0.
    \end{aligned}    
\end{equation}
Further, for the sake of simple calculation we put:
\begin{equation*}
	\label{}
	P=\left[\left[L+\left(\frac{N-1}{2}\right)\right]^2-\frac{1}{4}\right],
\end{equation*}
then Eq.~(\ref{15}) becomes:
\begin{equation}
	\label{16}
	\frac{d^2R}{dx^2}+\frac{2x}{x^2}\frac{dR}{dx}+\frac{2\mu}{x^4}\left[E-\frac{a_1}{x^2}-\frac{a_2}{x}+a_4x-\frac{P}{2\mu}x^2\right]R\left(x\right)=0.  
\end{equation}
Now, to make the equation linear in term of $x$ only, let us consider
\begin{equation}
	\label{17}
	x=y+\delta
\end{equation}
Here, $\delta$ is a free parameter, $\delta$=$\frac{1}{r_0}$. Here $r_0$ is defined as the characteristic radius of quarks, which is half of the separation distance of quarkonium system. 
Now by putting the value of $x$ from Eq.~(\ref{18}) and making binomial expansion, we get the values 
\begin{equation*}
	\label{}	
	\frac{a_1}{x^2}=\frac{a_1}{\delta}\left[3-\frac{2x}{\delta}\right].
\end{equation*}
\begin{equation*}
	\label{}	
	\frac{a_2}{x}=\frac{a_2}{\delta}\left[2-\frac{x}{\delta}\right]. 
\end{equation*}
Hence, with these values, Eq.~(\ref{17}) becomes:
\begin{equation}
	\begin{split}
	\label{18}
	\frac{\ d^2R}{dx^2}+\frac{2x}{x^2}\frac{dR}{dx}+\frac{1}{x^4}\left[2\mu\left(E-\frac{3a_1}{\delta}-\frac{2a_2}{\delta}\right)+\left(\frac{2a_1}{\delta^2}+\frac{a_2}{\delta^2}+\ \ a_4\right)x-Px^2\right]R\left(x\right)=0.
	\end{split}
\end{equation}
We can also write Eq.~(\ref{19})
\begin{equation}
	\label{19}
	\frac{d^2R}{dx^2}+\frac{2x}{x^2}\frac{dR}{dx}+\frac{1}{x^4}\left[-P_1+Q_1x-R_1x^2\right]R\left(x\right)=0,
\end{equation}
where the values of $P_1$, $Q_1$, $R_1$ is 
\begin{equation*}
	\label{}	
	P_1=-2\mu\left(E-\frac{3a_1}{\delta}-\frac{2a_2}{\delta}\right),
\end{equation*}
\begin{equation*}
	\label{}	
		Q_1=\left(\frac{2a_1}{\delta^2}+\frac{a_2}{\delta^2}+a_4\right),
\end{equation*}
\begin{equation*}
	\label{}	
	R_1=P. 
\end{equation*}
When we compare Eq.~(\ref{20}) with Eq.~(\ref{1}) we get
\begin{equation}
	\label{20}
{\bar\tau}(s)= 2x,\\
\sigma(s)= x^2,\\
{\sigma^2}(s)= x^4,\\
{\bar\sigma}(s)=-P_1+Q_1 x-R_1 x^2
\end{equation}
By putting these values in Eq.~(\ref{4}) we get the value of $\pi(x)$
\begin{equation}
	\label{21}
	\pi(x)=\pm\sqrt{(k+R_1 ) x^2-Q_1 x+P_1}.
\end{equation}

\begin{equation}
	\label{22}
	Q_1^2-4(k+R_1 ) P_1=0.
\end{equation}

\begin{equation}
	\label{23}
   k=\frac{Q_1^2}{4P_1}-R_1,
\end{equation}
Substituting 'k' in Eq.~(\ref{21}) we get
\begin{equation}
	\label{24}
	\pi(x)=±\frac{(Q_1 x-2P_1)}{\sqrt{4P_1}}.
\end{equation}
For unique solution, we only take positive value of  $\pi(x)$. On differentiating Eq.~(\ref{24}) we get,
\begin{equation}
	\label{25}
	\pi\prime (x)=\frac{Q_1}{\sqrt{4P_1}},
\end{equation}
Now putting the value of $\pi(x)$ from Eq.~(\ref{24}) and ${\bar\tau}$(s) from Eq.~(\ref{20}), we get the value of $\tau$(s) :
\begin{equation}
	\label{26}
	\tau(x)=2x+\frac{(Q_1 x-2P_1)}{P_1}.
\end{equation}
Differentiating Eq.~(\ref{26}) w.r.t. x we get:
\begin{equation}
	\label{27}
	\tau\prime(x)=2+\frac{Q_1}{\sqrt{P_1}}.
\end{equation}
On substituting the value of $\tau'$(s) from Eq.~(\ref{27}) and $\sigma''$(s) from Eq.~(\ref{20}) in Eq.~(\ref{6}), we get
\begin{equation}
	\label{28}
	\frac{Q_1^2}{4P_1}-R_1+\frac{Q_1}{\sqrt{4P_1}}=-n\left(2+\frac{Q_1}{\sqrt{P_1}}\right)-n^2-n. 
\end{equation}
Solving Eq.~(\ref{28}), we obtained the general expression of the binding energy (B.E) for N-Dimensions:
\begin{equation}
	\label{34}
	B.E=\frac{a_2}{\delta}(2-m_D )-\frac{2\mu\left[\frac{a_2}{\delta^2}\left(1-\frac{2m_D}{3}\right)+\alpha\right]^2}{\left[(2n+1)+\sqrt{1+4\left[\left(L+\frac{N-2}{2}\right)^2-\frac{1}{4}\right]}\right]^2}.
\end{equation}
Using this general expression, we obtained the variation of the binding energy of heavy quarkonium for different dimensionality and magnetic fields, as shown in Fig.~\ref{fig.1}, Fig.~\ref{fig.2} and Fig.~\ref{fig.3}:

\begin{figure*}[ht]
	\vspace{0mm}   
	\includegraphics[height=6cm,width=6cm]{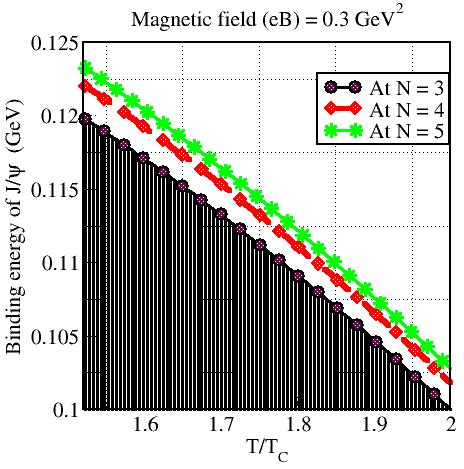}
	\hspace{0mm}
	\includegraphics[height=6cm,width=6cm]{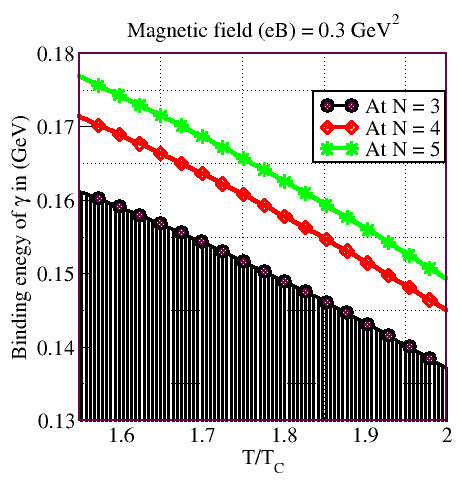}
	\caption{Shows the variation of binding energy of $J/\psi$ in left panel and for $\Upsilon$ in right panel for different values of dimensionality (N = 3, 4, 5) at fixed value of magnetic fields at $eB=0.3 GeV^2$.}
	\label{fig.1}
	\vspace{1mm} 
\end{figure*}


\begin{figure*}[ht]
	\vspace{0mm}   
	\includegraphics[height=6cm,width=6cm]{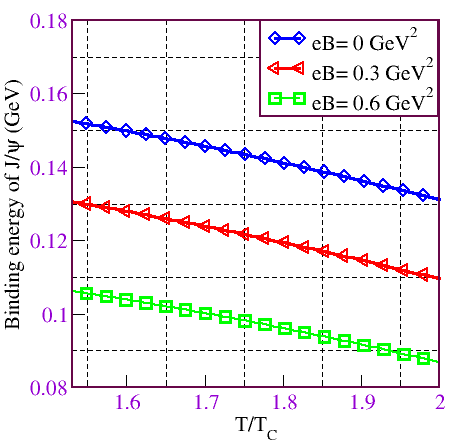}
	\hspace{0mm}
	\includegraphics[height=6cm,width=6cm]{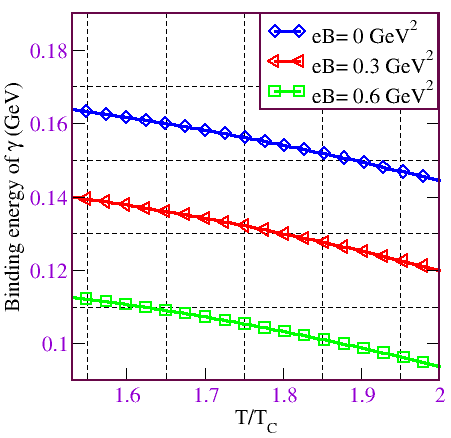}
	\caption{Shows the variation of binding energy of $J/\psi$ in left panel and for $\Upsilon$ in right panel with different magnetic fields at $N_f=3$.}
	\label{fig.2}
	\vspace{1mm} 
\end{figure*}
\begin{figure*} 
	\vspace{0mm}   
	\includegraphics[height=6cm,width=6cm]{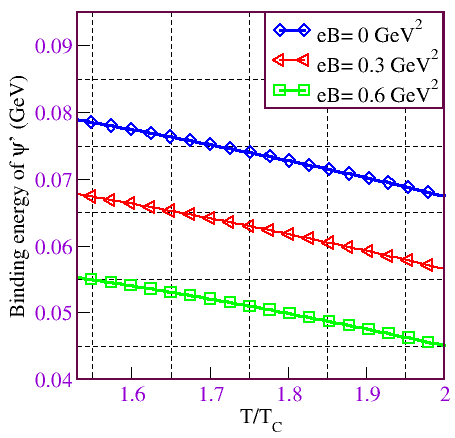}
	\hspace{1mm}
	\includegraphics[height=6cm,width=6cm]{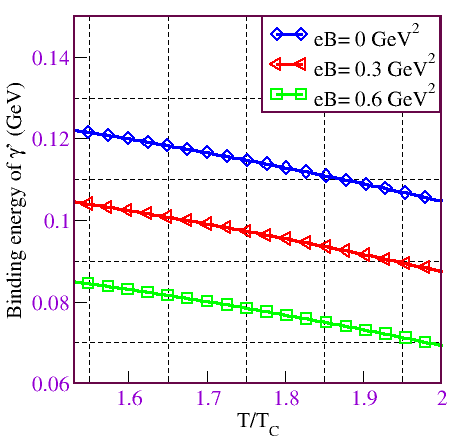}
	\caption{Shows the variation of binding energy of $\psi'$ in left panel and for $\Upsilon'$ in right panel with different magnetic fields at $N_f=3$.}
	\label{fig.3}
	\vspace{1mm} 
\end{figure*}

\subsection{Mass spectra}
The mass spectra of quarkonium states have been calculated using the following ansatz: 
\begin{equation}
	\label{36}
	Mass Spectra=2m_Q+ B.E,
\end{equation}
\begin{equation}
	\label{36}
	Mass Spectra=2m_Q+\frac{a_2}{\delta}(2-m_D )-\frac{2\mu\left[\frac{a_2}{\delta^2}\left(1-\frac{2m_D}{3}\right)+\alpha\right]^2}{\left[(2n+1)+\sqrt{1+4\left[\left(L+\frac{N-2}{2}\right)^2-\frac{1}{4}\right]}\right]^2}. 
\end{equation}

The variation of Mass spectra of heavy quarkonium at different values magnetic fields can be seen in Fig.~\ref{fig.4} and Fig.~\ref{fig.5}:

\begin{figure*}
	\vspace{0mm}   
	\includegraphics[height=6cm,width=6cm]{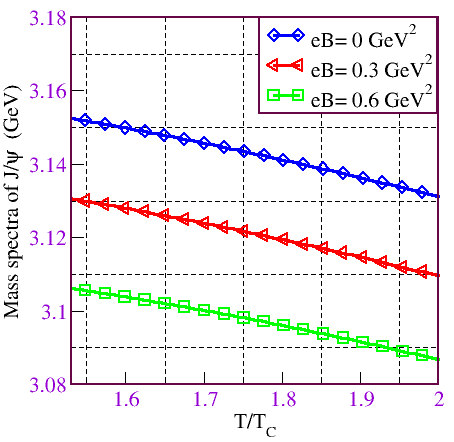}
	\hspace{1mm}
	\includegraphics[height=6cm,width=6cm]{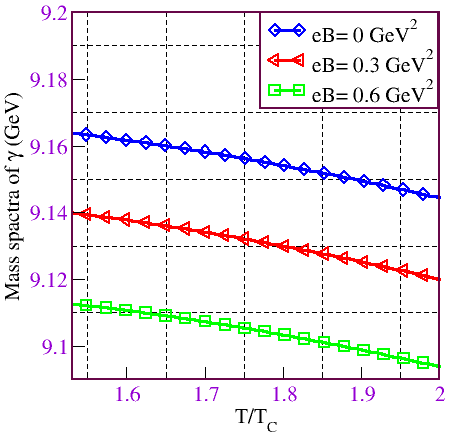}
	\caption{Shows the variation of Mass spectra of $J/\psi$ in left panel and for $\Upsilon$ in right panel with different magnetic fields at $N_f=3$.}
	\label{fig.4}
	\vspace{1mm} 
\end{figure*}
\begin{figure*}
	\vspace{0mm}   
	\includegraphics[height=6cm,width=6cm]{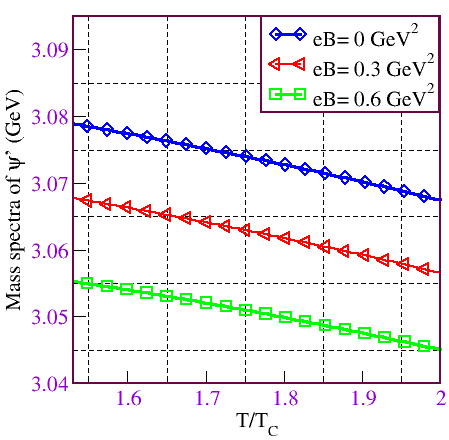}
	\hspace{1mm}
	\includegraphics[height=6cm,width=6cm]{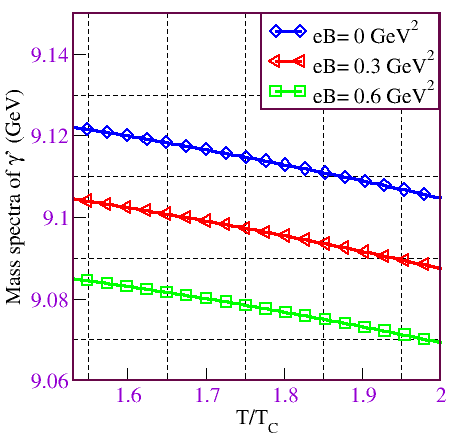}
	\caption{Shows the variation of Mass spectra of $\psi'$ in left panel and for $\Upsilon'$ in right panel with different magnetic fields at $N_f=3$.}
	\label{fig.5}
	\vspace{1mm} 
\end{figure*}

\section{Results and Summary} \label{sec3}
In the present work, we examine the properties of Charmonium and Bottomonium by solving Schrodinger equation using NU method in N-dimensional space. Here we have used medium modified form of Cornell potential (Eq.~(\ref{7})), the effect of magnetic field incorporated through the quasi particle Debye mass. We obtained the expression of energy eigenvalue in N-dimensional space, from which we study the quarkonia characteristics (i.e. binding energy and mass spectra). $\delta$ is a parameter whose numerical value we have determined with the reciprocal of characteristic radius~\cite{51a}. The values of $\delta$ will be different for $J/\psi=0.7892$ GeV, for $\Upsilon=1.4094$ GeV and for $\psi'=0.4384$ GeV, for $\Upsilon'=0.7074$ GeV. The heavy quark masses for $J/\Psi= 1.5$ GeV and $\Upsilon= 4.5$ GeV to calculate the quarkonium characteristics are taken from~\cite{53}. Fig.~\ref{fig.1} shows the binding energy of $J/\Psi$ (left panel) and $\Upsilon$ (right panel) with $T/T_C$ for different dimensionality number (N = 3, 4, 5) at fixed value of magnetic field (eB = 0.3 $GeV^2$). The binding energy decreases with increasing temperature, as demonstrated in Fig.~\ref{fig.1} Notably, the Fig.~\ref{fig.1} indicates that as the dimensionality number increases from 3 to 5, the initial binding energy values are higher, although the general trend of decreasing binding energy with temperature remains consistent. This suggests that higher dimensionality numbers result in stronger initial binding energies, but they still experience a decline as temperature rises.
Fig.~\ref{fig.2} shows the variation of binding energy of $J/\Psi$ (left panel) and $\Upsilon$ (right panel) and Fig.~\ref{fig.3} shows the variation of the binding energy of $\Psi\prime$ (left panel) and $\Upsilon\prime$ (right panel) with $T/T_C$ for different values of magnetic field (eB = 0, 0.3, 0.6 $GeV^2$) at fixed values of temperature (T= 0.3 GeV) and dimensionality (N=4). From both the Fig.~\ref{fig.2} and Fig.~\ref{fig.3}, we deduce that binding energy decreases with increasing values of magnetic field.
The variation of mass spectra with temperature  has been shown in Fig.~\ref{fig.4} and Fig.~\ref{fig.5} respectively. The left panel of Fig.~\ref{fig.4} and Fig.~\ref{fig.5} shows variation of mass spectra of Charmonium (i.e. $J/\Psi$  and $\Psi\prime$ ), whereas the right panel of Fig.~\ref{fig.4} and Fig.~\ref{fig.5} shows the variation of mass spectra for the Bottomonium i.e. ($\Upsilon$ and $\Upsilon\prime$)  at different magnetic field respectively. It is seen from Fig. 4 and Fig. 5 that, if we increase the value of magnetic field, then the mass spectra of both the quarkonium states decrease. Table~1 illustrates the significant influence of magnetic fields on the mass spectra of quarkonium states. By comparing our current results with previously published data (both theoretical and experimental), we find a reasonably good alignment with those findings.


\begin{table}[ph]
	\tbl{Tabular representation of Mass spectra  of different quarkonium states in (GeV) for different magnetic field.}
	{\begin{tabular}{@{}ccccccc@{}} \toprule
			$\mbox{\boldmath$States\downarrow$}$ & $\mbox{eB= 0 $GeV^2$}$ & $\mbox{eB= 0.3 $GeV^2$}$ & $\mbox{eB= 0.6 $GeV^2$}$ & Ref. \citen{44} & Ref. \citen{48} & Ref. \citen{52}  \\
			 \colrule
			$\mbox{\boldmath$J/\psi$}$ & 3.15 & 3.13 & 3.108 & 3.078 & 3.360 & 3.096 \\
			$\mbox{\boldmath$\psi'$}$ & 3.078 & 3.05 & 3.03 & 4.187 & 3.698 & 3.686 \\
			$\mbox{\boldmath$\Upsilon$}$ & 9.162 & 9.142 & 9.112 & 9.510 & 9.951 & 9.460 \\
			$\mbox{\boldmath$\Upsilon'$}$ & 9.12 & 9.11 & 9.07 & 10.62 & 10.03 & 10.023 \\ \botrule
		\end{tabular} \label{t1}}
\end{table}


	\markright{Bibliography}
{}

\end{document}